\documentclass[preprint,12pt]{elsarticle}

\usepackage{amssymb}
\usepackage{listings}
\usepackage{xcolor}
\usepackage{amsmath}
\usepackage{bm}
\usepackage{hyperref}

\definecolor{codegreen}{rgb}{0,0.6,0}
\definecolor{codegray}{rgb}{0.5,0.5,0.5}
\definecolor{codepurple}{rgb}{0.58,0,0.82}
\definecolor{backcolour}{rgb}{0.95,0.95,0.92}

\lstdefinestyle{mystyle}{
  backgroundcolor=\color{backcolour}, commentstyle=\color{codegreen},
  keywordstyle=\color{magenta},
  numberstyle=\tiny\color{codegray},
  stringstyle=\color{codepurple},
  basicstyle=\ttfamily\scriptsize,
  frame=shadowbox,
  language=Python,
  morekeywords={True, False},  
  breakatwhitespace=false,         
  breaklines=true,                 
  captionpos=b,                    
  keepspaces=true,                 
  numbers=left,                    
  numbersep=5pt,                  
  showspaces=false,                
  showstringspaces=false,
  showtabs=false,                  
  tabsize=2
}
\usepackage{subcaption}

\usepackage{float}
\restylefloat{table}

\journal{SoftwareX}

\makeatletter
\def\ps@pprintTitle{%
 \def\@oddfoot{\parbox{\textwidth}{\em\sffamily\footnotesize \copyright{} 2024 The Authors. This manuscript version is made available under the CC-BY 4.0 license \url{https://creativecommons.org/licenses/by/4.0/}. Original DOI: \url{https://doi.org/10.1016/j.softx.2024.101722}}}%
 \def\@evenfoot{\em\small Preprint \hfill \today}%
}
\makeatother

\begin{document}

\begin{frontmatter}

\title{JaxSGMC: Modular stochastic gradient MCMC in JAX}

\author[tum]{Stephan Thaler \fnref{first}}
\author[tum]{Paul Fuchs \fnref{first}}
\author[tum]{Ana Cukarska}
\author[tum]{Julija Zavadlav \fnref{mdsi}}

\address[tum]{Multiscale Modeling of Fluid Materials, Department of Engineering Physics and Computation,
TUM School of Engineering and Design, Technical University of Munich, Germany}
\address[mdsi]{Munich Data Science Institute, Technical University of Munich, Germany}

\fntext[first]{contributed equally}
\ead{julija.zavadlav@tum.de}

\begin{abstract}
We present \textit{JaxSGMC}, an application-agnostic library for stochastic gradient Markov chain Monte Carlo (SG-MCMC) in JAX. SG-MCMC schemes are uncertainty quantification (UQ) methods that scale to large datasets and high-dimensional models, enabling trustworthy neural network predictions via Bayesian deep learning. \textit{JaxSGMC} implements several state-of-the-art SG-MCMC samplers to promote UQ in deep learning by reducing the barriers of entry for switching from stochastic optimization to SG-MCMC sampling. Additionally, {\textit{JaxSGMC}} allows users to build custom samplers from standard SG-MCMC building blocks. 
Due to this modular structure, we anticipate that \textit{JaxSGMC} will accelerate research into novel SG-MCMC schemes and facilitate their application across a broad range of domains.

\end{abstract}


\end{frontmatter}

\section*{Code Metadata}

\begin{table}[H]
\begin{tabular}{|l|p{6.5cm}|p{6.5cm}|}
\hline
\textbf{Nr.} & \textbf{Code metadata description} & \textbf{Please fill in this column} \\
\hline
C1 & Current code version & v0.1.3 \\
\hline
C2 & Permanent link to code/repository used for this code version & \href{https://github.com/tummfm/jax-sgmc}{https://github.com/tummfm/jax-sgmc} \\
\hline
C3 & Code Ocean compute capsule & None\\
\hline
C4 & Legal Code License   & Apache-2.0 \\
\hline
C5 & Code versioning system used & git \\
\hline
C6 & Software code languages, tools, and services used & Python \\
\hline
C7 & Compilation requirements, operating environments \& dependencies & JAX\\
\hline
C8 & If available Link to developer documentation/manual & \href{https://jax-sgmc.readthedocs.io}{https://jax-sgmc.readthedocs.io} \\
\hline
C9 & Support email for questions & stephan.thaler@tum.de\\
\hline
\end{tabular}
\caption{Code metadata}
\end{table}

\section{Motivation and significance}

Deep learning models have seen enormous success in many scientific fields over the last decade, including disciplines as diverse as natural language processing \cite{Devlin2018}, autonomous driving \cite{Grigorescu2020}, health care \cite{Miotto2018} and physics-based modeling \cite{Raissi2019, Noe2020, Thaler_2021}. However, neural networks (NNs) are data-driven black-box models -- their predictions can be inaccurate when applied outside their training distribution \cite{Tossou2023,Arakelyan2023}. Uncertainty Quantification (UQ) provides a means to evaluate the trustworthiness of predictions, which is imperative for applying NNs in practice, in particular for safety-critical applications.

UQ approaches for neural networks include Bootstrapping \cite{Efron1994}, conformal inference \cite{Lei2018}, Deep Ensembles \cite{Lakshminarayanan2017} as well as Bayesian methods \cite{Neal2011, Welling2011, Graves2011}.
Bayesian statistics provides the mathematical foundation of Bayesian UQ, but classical Bayesian methods based on Markov chain Monte Carlo (MCMC) \cite{Neal2011, Hoffman2014} are intractable for computationally expensive NNs and large datasets \cite{Welling2011}. Stochastic gradient (SG) MCMC schemes \cite{Welling2011, Chen2014, Li2016, Nemeth2021, Lamb2020} circumvent the need for a full evaluation of the likelihood per parameter update of classical MCMC by leveraging a stochastic estimate of the gradient of the likelihood over a mini-batch of data. This results in a large computational speed-up, enabling Bayesian deep learning.

The landscape of UQ libraries is fragmented: There are domain-dependent libraries such as NeuralUQ \cite{Zou2022} on the one hand and domain-independent libraries on the other hand.
TensorFlow Probability \cite{Dillon2017} and Pyro \cite{Bingham2019} are the most popular domain-independent UQ libraries for Tensorflow and PyTorch, respectively. Both focus on classical Hamiltonian Monte Carlo \cite{Neal2011} schemes and Stochastic Variational Inference \cite{Hoffman2013}, while Stochastic Gradient Langevin Dynamics (SGLD) is the only implemented SG-MCMC sampler.
Thus, dedicated SG-MCMC libraries have been developed for Tensorflow \cite{Baker2019}, Theano \cite{Gupta2016} and JAX \cite{Coullon2022}. However, the structure of these libraries currently does not allow for recently proposed SG-MCMC building blocks such as parallel tempering \cite{Deng2020} and amortized Metropolis Hastings (MH) acceptance steps \cite{Zhang2020, Garriga2020}. Hence, many newly developed SG-MCMC samplers are published as stand-alone code \cite{Gallego2018, Deng2020, Zhang2020} and do not take advantage of these existing libraries, which slows adoption of novel SG-MCMC samplers in practice.

In this work, we introduce the domain-independent \textit{JaxSGMC} library. \textit{JaxSGMC} implements several state-of-the-art SG-MCMC samplers such as replica exchange SG-MCMC \cite{Deng2020} and AMAGOLD \cite{Zhang2020}. The implemented SG-MCMC schemes follow a common application programming interface (API), which simplifies switching between samplers and reduces the barriers of entry to UQ for practitioners.
The SG-MCMC samplers are designed in a modular fashion, which allows re-using standard SG-MCMC building blocks.
Additionally, the samplers can be compiled end-to-end just-in-time (jit), which improves their computational efficiency.

\section{Software description}

\subsection{Bayesian Modeling}

A model consists of an architecture $\mathcal M$ and parameters $\bm \theta$.
In deep learning, these models are NNs, such as ResNet \cite{He2016}, with millions of parameters.
The frequentist machine learning (ML) approach selects a good model via stochastic optimization of a loss function to obtain an optimal set of parameters $\bar{\bm \theta}$ that best fits a dataset $\mathcal D$ (e.g. CIFAR-10 \cite{cifar10}).
The selected $\bar{\bm \theta}$ critically determines the model performance and reliability.

In contrast, the Bayesian approach studies the posterior predictive distribution
\begin{equation}
  p(\mathrm y\vert \mathbf x, \mathcal D, \mathcal M)  = \int p(\mathrm y \vert \mathbf x, \bm \theta, \mathcal M)p(\bm\theta \vert \mathcal D, \mathcal M)\mathrm{d}\bm\theta \ , \label{eq:unvertainty_aware}
\end{equation}
which encodes the uncertainty in the model prediction $y$ given an input $\mathbf x$.
Instead of betting on a single parameter set $\bar{\bm \theta}$, eq. \eqref{eq:unvertainty_aware}
considers an infinite number of models weighted according to their agreement with the dataset given by the posterior distribution $p(\bm{\theta}|\mathcal{D}, \mathcal M)$.
This integral is analytically intractable,
but can be estimated via Monte Carlo integration employing a finite number of models
\begin{equation}
  p(\mathrm y\vert \mathbf x, \mathcal D, \mathcal M) \approx  \frac{1}{N_\text{models}} \sum_{i=1}^{N_\text{models}} p(\mathrm y\vert \mathbf x,\bm \theta_{i}, \mathcal M); \quad \  \bm \theta_{i} \sim p(\bm{\theta}\vert\mathcal{D}, \mathcal M)\,
  \label{eq:Monte_Carlo}
\end{equation} drawn from the posterior distribution. Bayes formula relates the posterior 
\begin{equation}
  p(\bm \theta\vert\mathcal{D}, \mathcal M) = \frac{p(\mathcal{D}\vert\bm \theta, \mathcal M)p(\bm \theta \vert \mathcal M)}{p(\mathcal{D} \vert \mathcal M)} \propto \exp\left(-\mathcal U(\bm\theta) \right)
  \label{eq:potential_energy_relation}
\end{equation}
to the likelihood $p(\mathcal D \vert \bm \theta, \mathcal M)$, prior $p(\bm \theta \vert \mathcal M)$ and model evidence $p(\mathcal D \vert \mathcal M)$, which normalizes the distribution.
The likelihood is the probability that a model could have generated the data. Complementary, the prior encodes beliefs about the model, independent of the data. Likelihood and prior are closely connected to loss functions and regularization techniques commonly used in the frequentist ML approach.
However, no known method exists that can generate independent samples from arbitrary distributions.

Instead, modern MCMC algorithms propose sequences of samples by simulating physical processes such as Hamiltonian or Langevin dynamics, which are driven by the gradient of the potential $\mathcal U(\bm \theta)$ (eq. \eqref{eq:potential_energy_relation}).
An additional MH step accepts or rejects each proposal such that the equilibrium distribution of the Markov chain agrees with the posterior distribution \cite{Hastings1970}. Extending gradient-based MCMC approaches to big data applications is computationally infeasible due to the dependence of the gradient $\nabla \mathcal U(\bm \theta)$ on all data points, which needs to be computed at each timestep of the simulation.

Similar to stochastic gradient descent (SGD) schemes, SG-MCMC methods resort to a noisy estimate of the potential
\begin{equation}
  \mathcal{U}(\bm{\theta}) \approx -\frac{N}{n}\sum\limits_{i=1}^{n} \log p(\mathrm y_i \vert \mathbf x_i, \bm \theta, \mathcal M) - \log p(\bm\theta \vert \mathcal M)
  \label{eq:potential_energy_function}
\end{equation}
based on a random mini-batch of $n$ data points \cite{Welling2011}.
However, these approximate dynamics bias the equilibrium distribution and render the conventional MH corrections inapplicable \cite{Welling2011, Garriga2020}. Nevertheless, classical SG-MCMC schemes achieve an asymptotically correct equilibrium distribution by adequately annealing the simulation timestep $\Delta t \rightarrow 0$ and adding the right amount of noise to the stochastic gradient \cite{Welling2011, ma2015}.

More recent SG-MCMC schemes offer enhanced MH steps to sample from the unbiased distribution at finite $\Delta t$. These MH steps accept or reject multiple consecutive proposals, while requiring a full potential evaluation only once \cite{Zhang2020, Garriga2020}.
Other schemes aim to improve the mixing behavior of the Markov chain on highly curved NN posteriors by extending stochastic optimization algorithms, such as Adam and preconditioned SGD, to SG-MCMC \cite{KimSongLiang2022, Li2016}. Additionally, tempered and multi-chain algorithms enhance exploratory capabilities to address high posterior multi-modality, e.g. by cyclically annealing temperature and timestep size \cite{Zhang2019cyc} or swapping samples between tempered and non-tempered chains \cite{Deng2020}. 

\subsection{Software Architecture}

\begin{figure}[tb]
    \centering
    \includegraphics[width=\textwidth]{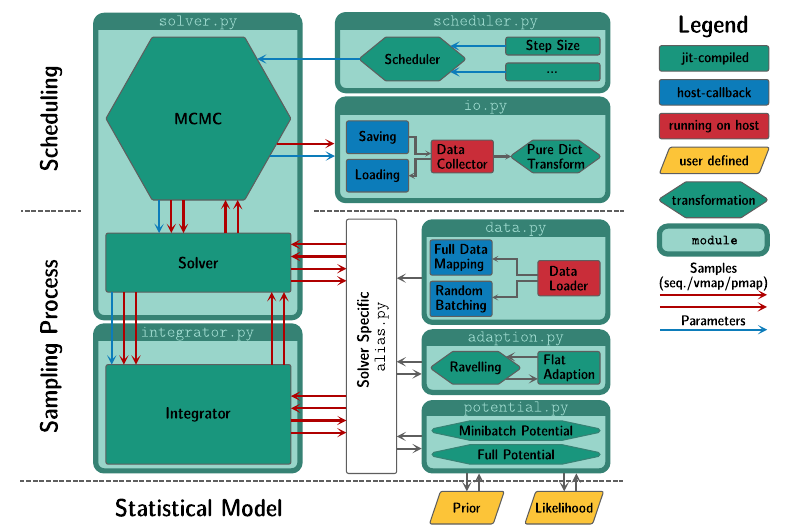}
    \caption{Software architecture of \textit{JaxSGMC}. The modules are ordered hierarchically from left to right, and the specificity of a module to the machine learning problem increases from top to bottom.}
    \label{fig:jax-sgmc-structure}
\end{figure}

Fig.~\ref{fig:jax-sgmc-structure} visualizes the relationship between the software modules in \textit{JaxSGMC}.
Each module contains algorithmic building blocks that form a SG-MCMC sampler.
Concisely, an SG-MCMC sampler repeats the following steps: (a) Retrieve the current process parameters ($\Delta t$, temperature, \ldots),  (b) simulate the process via the potential $\mathcal U$ starting from the current sample $\bm \theta_i$, (c) process the obtained proposal and (d) save the new sample $\bm \theta_{i+1}$. The specific implementation of these steps defines the SG-MCMC sampler.

Each sampler builds on a user-provided model, which can be any JAX-transformable function, e.g. a NN. The model is part of the log-likelihood, which -- together with the log-prior -- forms the potential $\mathcal U(\bm \theta)$ (eq. \eqref{eq:potential_energy_function}). $\mathcal U(\bm \theta)$ represents the statistical model and links the model and the sampler. Although sometimes plainly referred to as likelihood and prior, all computations rely on log-transformed values for computational reasons.
Due to the (assumed) independence of observations in the dataset, the user-provided log-likelihood function computes the log-probability $p(\mathrm y_k \vert \mathbf x_k, \bm \theta_i, \mathcal M)$ of the current model $\bm \theta_i$ for a single observation $(\mathbf x_k, \mathrm y_k) \in \mathcal D$ drawn from the dataset. 
The functions in the \texttt{potential.py} module then apply the log-probability to mini-batches of data to compute $\mathcal{U}(\bm{\theta}_i)$ or its stochastic estimate (eq.~\eqref{eq:potential_energy_function}).
By designing \textit{JaxSGMC} in a functional programming style, we leverage JAX's function transformations and automatic differentiation throughout the framework to perform these batched evaluations of the log-likelihood,
to calculate the (stochastic) gradients $\nabla \mathcal U(\bm \theta)$ and run multiple Markov chains.

\textit{JaxSGMC} focuses on the big data case, where storing the whole dataset on the device (GPU) is inefficient or impossible due to memory limitations.
Yet, functions relying on data should support JAX transformations. To this end, the \texttt{data.py} module offers an API around the host-callback module of JAX to efficiently insert data
from so-called DataLoaders into jit-compiled computations. 
By providing different DataLoaders, we support multiple mini-batch assembly methods as well as straightforward data integration from different sources.
Similarly, already a small number of parameter samples $\bm \theta_i$ of deep NNs can fill up the device memory. Accordingly, we designed the \texttt{io.py} module analogously to the \texttt{data.py} module to efficiently store the gathered samples. Hence, \textit{JaxSGMC} enables collecting many samples from within the jit-compiled SG-MCMC algorithm and saving them to different file formats.

The \texttt{adaption.py} module, which supports the adaption of process quantities to the learning problem, together with the \texttt{data.py} and \texttt{potential.py} modules, provide all relevant components for the core of a SG-MCMC algorithm. 
This core lies in the sampling section simulating the physical process (step (b)) and processing the proposals (step (c)). 
In line with the modularity design principle, \textit{JaxSGMC} subsequently separates the former into the \texttt{integrator.py} module and the latter into the \texttt{solver.py} module.

The scheduling part of the algorithm builds on top of the sampling process (fig. \ref{fig:jax-sgmc-structure}).
The scheduling section includes boilerplate code, which provides a general entry point to run a constructed algorithm. In particular, it interfaces the schedules of the process parameters (step (a)) in \texttt{scheduler.py} with the Markov chain (steps (b) and (c)) and saves the current solver state or the collected samples (step (d)). 
Typically, the schedules are static and thus independent of the sampling process, but \textit{JaxSGMC} also supports a feedback loop to enable adaptive step-size schemes.

\subsection{Software Functionalities}

We implemented two API levels: First, we built a high-level interface to popular SG-MCMC samplers in \texttt{alias.py}, including (preconditioned) SGLD \cite{Welling2011, Li2016}, stochastic gradient Hamiltonian Monte Carlo (SGHMC) \cite{Chen2014}, replica exchange SG-MCMC (reSGLD) \cite{Deng2020}, AMAGOLD \cite{Zhang2020}, and stochastic gradient guided Monte Carlo (SGGMC) \cite{Garriga2020}.
This interface aims to enable users with an existing dataset and a JAX model to easily switch from stochastic optimizers to SG-MCMC samplers, while still providing flexibility in the dataset format and stochastic potential evaluation strategy.

A second API level is inspired by the stochastic optimization library Optax \cite{Deepmind2020}. It allows more advanced users to combine SG-MCMC building blocks to create custom samplers tailored to the individual problem. Table~\ref{tab:building_blocks} gives an overview of the currently implemented algorithmic building blocks and supported data formats.
In addition, the implemented DataLoaders enable end-to-end jit-compilation of learning algorithms for maximum computational efficiency, even beyond the scope of SG-MCMC.

\begin{table}
    \begin{tabular}{|c|p{.72\textwidth}|}\hline
        \textbf{Module} & \multicolumn{1}{c|}{\textbf{Content}}\\\hline
         \texttt{adaption.py} & \textbf{Algorithms:} RMSProp \cite{Tieleman2012}, online covariance estimation, Fisher Information estimation \cite{Sungjin2012} \\\hline
         \texttt{integrator.py} & \textbf{Process simulators:} OBABO \cite{Garriga2020}, time-reversible leapfrog \cite{Zhang2020}, leapfrog with friction \cite{Chen2014}, Langevin diffusion \cite{Welling2011} \\\hline
         \texttt{scheduler.py} & \textbf{Schedules:}  adaptive step size \cite{Hoffman2014}, polynomial step size with optimal decay \cite{Teh2016}, constant temperature, initial burn-in, random thinning\\\hline
         \texttt{potential.py} & \textbf{Potentials:} stochastic potential, true potential \\\hline
         \texttt{data.py} & \textbf{Data sources:} numpy/JAX arrays, TensorFlow dataset, HDF5 \newline \textbf{Data batching:} mini-batching (drawing / shuffling / shuffling in epochs), batched mapping across full dataset\\\hline
         \texttt{io.py} & \textbf{Output formats:} numpy/JAX arrays, JSON, HDF5 \\\hline
         \texttt{solvers.py} & \multicolumn{1}{c|}{\textit{Lower-level interface to solvers in} \texttt{alias.py}} \\\hline
    \end{tabular}
    \caption{Overview of algorithmic building blocks in each \textit{JaxSGMC module.}}
    \label{tab:building_blocks}
\end{table}

\section{Illustrative Examples}
We provide two ML examples, each illustrating a different use-case of \textit{JaxSGMC}: building a custom sampler in a linear regression problem and using a pre-built sampler in an image classification problem. The interested reader may refer to the examples in the GitHub repository for more details.

\subsection{Linear Regression}
Many of the functionalities of \textit{JaxSGMC} can be introduced with a simple linear regression model.
Dataset arrays can be passed as keyword arguments to a DataLoader (\texttt{data.py}). The DataLoader stores the dataset on the host (CPU) and can orchestrate sending mini-batches to the device (e.g. GPU) as they are requested (listing \ref{listing:load_dataset}).

\begin{lstlisting}[language=Python, caption=Loading a dataset with \textit{JaxSGMC}., label={listing:load_dataset}]
from jax_sgmc.data.numpy_loader import NumpyDataLoader
from jax_sgmc.data import random_reference_data

data_loader = NumpyDataLoader(x=training_data_x, 
                              y=training_data_y)
                              
data_fn = random_reference_data(data_loader,
                                mb_size=batch_size,
                                cached_batches_count=100)
\end{lstlisting}
The next step is to define the linear model with weights $\mathbf{w}$, as well as log-likelihood and log-prior.
The log-likelihood follows from the assumption that the data includes Gaussian-distributed noise with mean 0 and a (learned) homoscedastic standard deviation $\sigma$.
The log-prior consists of an (improper) uniform distribution for $\mathbf{w}$ and an exponential distribution for the $\sigma$ parameter.
Log-likelihood and log-prior define the (mini-batch) potential (\texttt{potential.py}, eq. \eqref{eq:potential_energy_function}, listing \ref{listing:model}).
\begin{lstlisting}[language=Python, caption=Defining the stochastic potential function from the log-likelihood and log-prior., label={listing:model}]
from jax_sgmc import potential

def model(sample, observations):
    weights = sample["w"]
    predictors = observations["x"]
    return jnp.dot(predictors, weights)
    
def log_likelihood(sample, observations):
    sigma = jnp.exp(sample["log_sigma"])
    y = observations["y"]
    y_pred = model(sample, observations)
    return jax.scipy.stats.norm.logpdf(y - y_pred, loc=0, scale=sigma)

def log_prior(sample):
    return 1 / jnp.exp(sample["log_sigma"])

potential_fn = potential.minibatch_potential(prior=log_prior,
                                             likelihood=log_likelihood) 
\end{lstlisting}
The MemoryCollector (\texttt{io.py}) stores the sampled models in the host's working memory.
We implement the RMSProp \cite{Tieleman2012} preconditioned Stochastic Gradient Langevin Dynamics (pSGLD) method \cite{Li2016}, which
can be defined using RMSProp from the \texttt{adaption.py} module, a Langevin diffusion simulator (\texttt{integrator.py}) and a solver that accepts each sample unconditionally (\texttt{solver.py}).
The schedulers in the \texttt{scheduler.py} module operate independently from the solver and manage the stepsize, burn-in and thinning along the Markov chain.
The combination of these building blocks to obtain the pSGLD sampler is shown in listing \ref{listing:initialization}.

\begin{lstlisting}[language=Python, caption=Building the preconditioned Stochastic Gradient Langevin Dynamics sampler from its building blocks., label={listing:initialization}]
from jax_sgmc import io, adaption, integrator, solver
from jax_sgmc.scheduler import polynomial_step_size_first_last, initial_burn_in, random_thinning, init_scheduler

my_data_collector = io.MemoryCollector()
save_fn = io.save(data_collector=my_data_collector)

rms_prop_adaption = adaption.rms_prop()

ld_integrator = integrator.langevin_diffusion(potential_fn=potential_fn,
                                              batch_fn=data_fn,
                                              adaption=rms_prop_adaption)
                                              
rms_prop_solver = solver.sgmc(ld_integrator)


#Initialize the solver by providing initial values for the latent variables
init_sample = {"log_sigma": jnp.array(0.0), "w": jnp.zeros(N)}

init_state = rms_prop_solver[0](init_sample)

step_size_schedule = polynomial_step_size_first_last(first=0.05,
                                                     last=0.001,
                                                     gamma=0.33)
burn_in_schedule = initial_burn_in(2000)
thinning_schedule = random_thinning(step_size_schedule=step_size_schedule,
                                    burn_in_schedule=burn_in_schedule,
                                    selections=1000)

schedule = init_scheduler(step_size=step_size_schedule,
                          burn_in=burn_in_schedule,
                          thinning=thinning_schedule)

mcmc = solver.mcmc(solver=rms_prop_solver,
                   scheduler=schedule,
                   saving=save_fn)
\end{lstlisting}

Now the SG-MCMC sampling procedure can be performed.
Afterwards, the saved samples can be accessed for postprocessing (listing \ref{listing:run}).
\begin{lstlisting}[language=Python, caption=Sampling and accessing the results., label={listing:run}]
# Take the result of the first chain
results = mcmc(init_state, iterations=10000)[0]

print(f"Collected {results['sample_count']} samples")

sigma_rms = onp.exp(results["samples"]["variables"]["log_sigma"]
w_rms = results["samples"]["variables"]["w"]
\end{lstlisting}

\begin{figure}[htbp]
    \centering
    \includegraphics[scale=0.5]{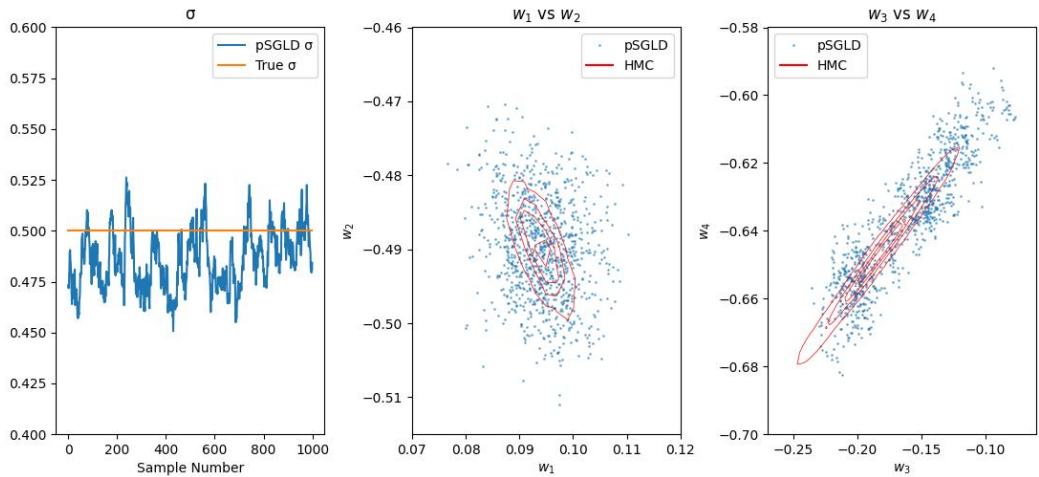}
    \caption{Left: Sampled standard deviation $\sigma$ parameters (blue) compared to the data-generating value (orange). Middle and right: First and second (Middle) and third and fourth (right) components of weights $w_i$ sampled with the pSGLD scheme (blue scatter plot) compared to contour plots of Gaussians obtained from the Hamiltonian Monte Carlo (HMC) method (red) implemented in NumPyro \cite{phan2019composable, Neal2011}.}
    \label{fig:linear_regression}
\end{figure}

We visualize the sampled parameters and compare the resulting distribution to a gold-standard Hamiltonian Monte Carlo scheme implemented in the NumPyro library \cite{phan2019composable, Neal2011} (fig. \ref{fig:linear_regression}).
The obtained distributions of both methods agree reasonably well, in line with expectations.

\subsection{Image Classification on CIFAR-10}

Next, we provide an example more typical for recent deep learning models. In particular, we consider an image classification task with the CIFAR-10 dataset \cite{cifar10}, which we split into a training, validation and test set containing 50000, 5000 and 5000 images, respectively. For the architecture of the NN, we use the 2.1 million parameter Haiku \cite{haiku2020github} implementation of MobileNet version 1 \cite{mobilenets} without batch normalization.
Given that the MobileNet architecture shows superior performance with larger images, we resized the images from 32x32 to 112x112 pixels using bilinear interpolation. 
The log-likelihood for a multiclass classification problem corresponds to the negative cross entropy. As prior distribution over NN weights and biases $\mathbf{w}$, we select a Gaussian centered at 0 with standard deviation of 10. With these components, the potential function can be defined (listing \ref{listing:mobilenet}).

\begin{lstlisting}[language=Python, caption={Creating a MobileNet version 1 \cite{mobilenets} NN model using Haiku \cite{haiku2020github} and defining log-likelihood, log-prior, and the (mini-batch) potential functions.}, label={listing:mobilenet}]
import haiku as hk, optax, tree_math
from jax import tree_map
from functools import partial
from jax_sgmc import potential

def init_mobilenet():
    @hk.transform
    def mobilenetv1(batch, is_training=True):
        images = batch["image"].astype(jnp.float32)
        mobilenet = hk.nets.MobileNetV1(num_classes=num_classes,
                                        use_bn=False)
        logits = mobilenet(images, is_training=is_training)
        return logits
    return mobilenetv1.init, mobilenetv1.apply
    
init_mobilenet, apply_mobilenet = init_mobilenet()
    
def log_likelihood(sample, observations):
    logits = apply_mobilenet(sample["w"], None, observations)
    log_likelihood = -optax.softmax_cross_entropy_with_integer_labels(
        logits, observations["label"])
    return log_likelihood
    
def log_gaussian_prior(sample):
    gaussian = partial(jscipy.stats.norm.logpdf, loc=0, scale=10)
    priors = tree_map(gaussian, sample["w"])
    return tree_math.Vector(priors).sum()
    
potential_fn = potential.minibatch_potential(prior=log_gaussian_prior,
                                             likelihood=log_likelihood,
                                             is_batched=True,
                                             strategy='vmap')
\end{lstlisting}

We use a pSGLD sampler with RMSProp preconditioner \cite{Li2016}, which can be set up with the ready-to-use sampler interface of the \texttt{alias.py} module (listing \ref{listing:sampler}). To initialize the sampler, the potential function, the DataLoader, and a set of hyperparameters need to be passed.
We cache 10 batches of data in the device memory and set the batch size to 256 images.
The learning rate is initially set to 0.001 and controlled by a polynomial step size scheduler.

\begin{lstlisting}[language=Python, caption={Defining a SG-MCMC sampler via the \texttt{alias.py} API with subsequent sampling.}, label={listing:sampler}]
from jax_sgmc import alias

sampler = alias.sgld(potential_fn=potential_fn,
                     data_loader=train_loader,
                     cache_size=cached_batches,
                     batch_size=batch_size,
                     first_step_size=lr_first,
                     last_step_size=lr_last,
                     burn_in=burn_in_size,
                     accepted_samples=accepted_samples,
                     rms_prop=True,
                     progress_bar=True)

results = sampler(sample, iterations=39000)
results = results[0]['samples']['variables']
\end{lstlisting}

We sample for 39000 iterations - corresponding to 200 epochs - and retain 20 NN models via random thinning after a burn-in period of 35100 iterations.
Interestingly, the cost of the whole pSGLD training of 200 epochs is the same order of magnitude as the cost of generating a single sample with the full-batch Hamiltonian Monte Carlo (HMC) method given that a single Hamiltonian integration step requires a full epoch of gradient computations and $\mathcal{O}(100)$ integration steps are needed for each parameter proposal (neglecting the burn-in period). This relation highlights the computational savings of SG-MCMC compared to classical full-batch MCMC methods.
The pSGLD training results in a training accuracy of 65.25\%, a validation accuracy of 55.72\% and a test accuracy of 57.32\%, which is evaluated via soft voting of the ensemble \cite{Kim2018AutomatedML}.
We validate this result by comparing it to a deterministic model with the same model architecture and hyperparameters, and find a comparable performance. 
Furthermore, the runtimes of the SG-MCMC sampling and the stochastic optimization are comparable.

\begin{figure}[htbp]
    \centering
        \begin{subfigure}[b]{\textwidth}   
        \centering 
        \includegraphics[scale=0.415]{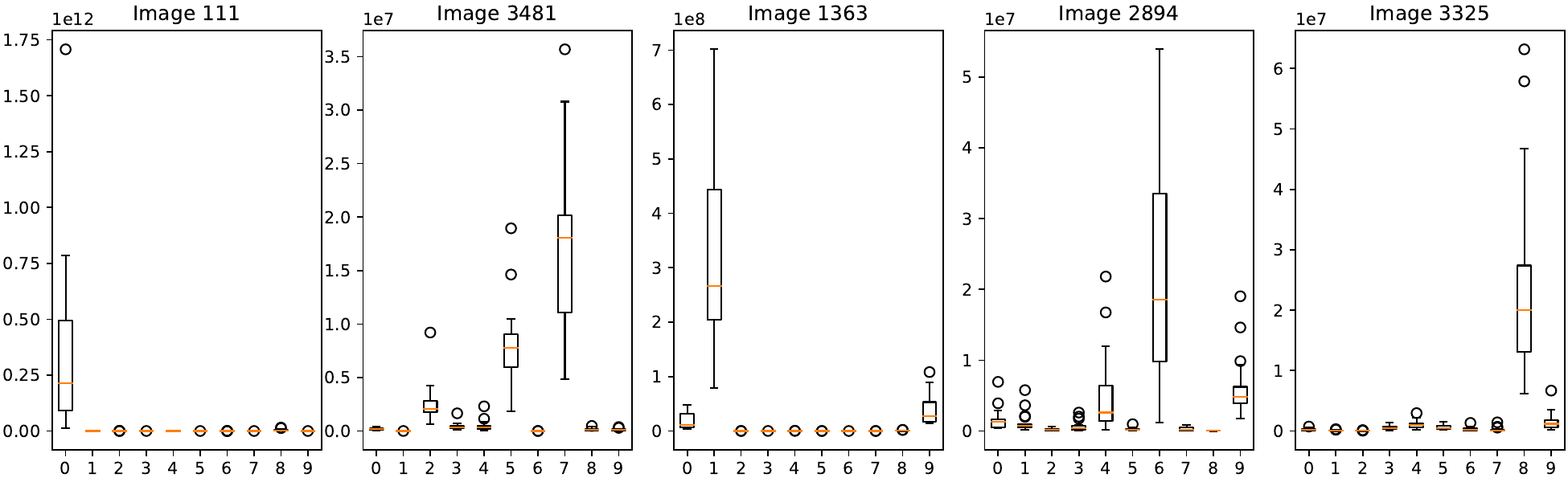}
        \caption{50000 images}
    \end{subfigure}
    \begin{subfigure}[b]{\textwidth}   
        \centering 
        \includegraphics[scale=0.415]{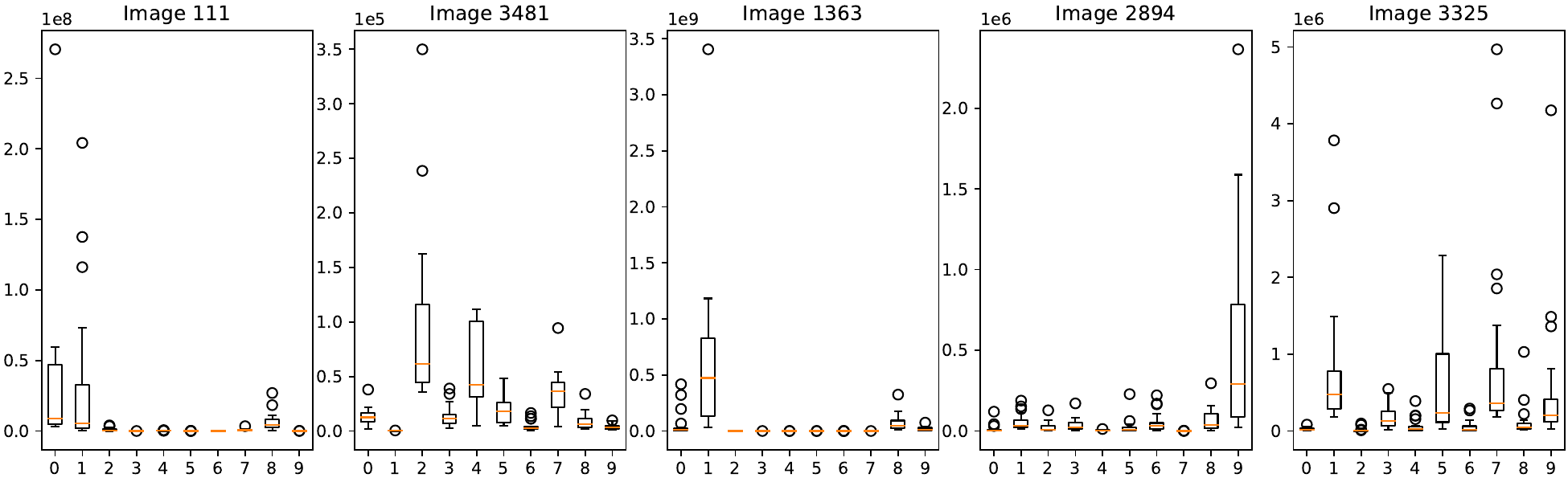}
        \caption{10000 images}
    \end{subfigure}
    \caption{Obtained distributions of logits for five randomly selected images from the testset visualized as box plots using (a) all 50000 training images and (b) a subset of 10000 training images. The true labels for the images in the order above are: 0, 7, 1, 6, 6.}
    \label{fig:predictions}
\end{figure}

An advantage of using SG-MCMC is that the distribution of predictions from the sampled models can readily be used for UQ.
As an example, we take five random images from the testset and visualize the distribution of the logits of each class in a box plot (fig. \ref{fig:predictions} (a)). In this example, the posterior predictive distribution is mostly concentrated on a single class -- the true class in the first 4 example images, but an incorrect class in example image 3325, resulting in an overconfident prediction. Given that the posterior variance is a function of the amount of training data, we train a new model for 200 epochs on a 10000 image training data subset. As expected, with less training data, the uncertainty of the predictions increases (fig. \ref{fig:predictions} (b), e.g. image 3325) and the soft voting test set accuracy decreases to 42.70\%.

Finally, we assess the quality of the uncertainty estimates from the model trained on the full training data by computing the testset accuracy for images, where the certainty of the prediction exceeds a specific threshold. We employ a hard voting-based \cite{Kim2018AutomatedML} estimate of the prediction certainty, i.e. the percentage of all sampled models that predicted the majority class.
As expected, the accuracy increases with increasing prediction certainty (table \ref{tab:accuracy}). However, for higher certainty thresholds, the obtained models are overconfident.
\begin{table}[htbp]
\begin{tabular}{ | c | c | c | c | c | c | c |}
\hline
 certainty & $\ge$ 50\% & $\ge$ 60\% & $\ge$ 70\% & $\ge$ 80\% & $\ge$ 90\% & = 100\%  \\ 
\hline
 validation accuracy & 58.89 & 61.72 & 64.61 & 67.60 & 71.64 & 77.95 \\
\hline
 test accuracy & 60.11 & 63.06 & 66.59 & 70.20 & 74.65 & 80.81 \\
\hline
\end{tabular}
\caption{\label{tab:accuracy}Accuracy depending on the certainty of the ensemble.}
\end{table}

In this example, we opted for pSGLD \cite{Li2016}, a comparatively simple SG-MCMC scheme. The accuracy and the quality of UQ could probably be increased by leveraging more advanced SG-MCMC components provided by \textit{JaxSGMC}, including running multiple Markov chains \cite{Thaler2022d}.
However, this is beyond the scope of this illustrative example.

\section{Impact and Conclusion}

Trustworthy predictions via uncertainty-aware ML \cite{Wang2020survey} and increasing data efficiency via active learning \cite{Ren2021} represent highly promising avenues in deep learning.
However, the effectiveness of these approaches critically depends on the quality of UQ estimates.
To this end, it is insufficient to sample only a single NN posterior mode \cite{Wilson2020}, e.g. when using Stochastic Variation Inference \cite{Hoffman2013} or Dropout Monte Carlo \cite{Gal2016}. While the popular Deep Ensemble \cite{Hansen90, Lakshminarayanan2017} scheme captures different posterior modes, it neglects the uncertainty contribution from the volume of the posterior.
In contrast, SG-MCMC, especially when using multiple Markov chains, can sample multiple modes as well as the volume of the posterior.

Despite this theoretical advantage, SG-MCMC schemes are still underused in practice, in part due to a lack of easy-to-use libraries that implement state-of-the-art SG-MCMC samplers. \textit{JaxSGMC} simplifies switching from stochastic optimization to Bayesian sampling by providing a common API (\texttt{alias.py}) for SG-MCMC samplers, which can replace stochastic optimizers \cite{Deepmind2020} without modifications to the JAX NN model.
Hence, \textit{JaxSGMC} makes recently developed SG-MCMC samplers available to a broader user base.
Additionally, by building custom samplers, the SG-MCMC schemes can be tailored to the ML problem at hand.

\textit{JaxSGMC} is an domain-independent library. As such, we selected classical ML benchmark problems for the examples presented in this paper, but the provided code can also be used for other applications such as physics-based modeling.
A recent example is the training of NN potentials \cite{Thaler2022}, where the jit-compatible DataLoaders of \textit{JaxSGMC} are used to improve computational performance and simplify implementation by integrating data loading into the jit-compiled parameter update function.
Furthermore, the pre-implemented SG-MCMC samplers of \textit{JaxSGMC} have been used to switch from stochastic optimization to Bayesian inference for cases of molecular modeling with classical \cite{Thaler2022b} and NN potentials \cite{Thaler2022d}. These studies have shown that pSGLD does not yet fully exploit the theoretical advantage of additional exploration of the posterior volume \cite{Thaler2022d}. Thus, further research into SG-MCMC samplers with more sophisticated posterior exploration capabilities is required, which can be accelerated with \textit{JaxSGMC}.
We envision that \textit{JaxSGMC} will promote uncertainty-aware ML and active learning applications in physical modeling and beyond.

\section{Conflict of Interest}
We wish to confirm that there are no known conflicts of interest associated with this publication and there has been no significant financial support for this work that could have influenced its outcome.

\section*{Acknowledgements}
S.T. acknowledges financial support from the Munich Data Science Institute Seed Fund.

Funded by the European Union. Views and opinions expressed are however those of the author(s) only and do not necessarily reflect those of the European Union or the
European Research Council Executive Agency. Neither the European Union nor the granting
authority can be held responsible for them. Funded by the European Research Council (ERC) StG under Grant No. 101077842—SupraModel.

\bibliographystyle{elsarticle-num} 
\bibliography{library}

\end{document}